# Calculation of Thermodynamic Equilibria with the Predictive Electrolyte Model COSMO-RS-ES: Improvements for Low Permittivity Systems

Simon Müller[a]*, Andrés González de Castilla[a], Christoph Taeschler[b], Andreas Klein[b], Irina Smirnova[a]

* corresponding author (simon.mueller@tuhh.de)

[a] *Institute of Thermal Separation Processes, Hamburg University of Technology, Hamburg, Germany*

[b] *Lonza AG, Visp, Switzerland*

**Abstract**

The predictive electrolyte model COSMO-RS-ES is refined to improve the description of systems at 25°C in which strong ion pairing is expected due to a low static permittivity of the liquid phase. Furthermore, the short-range ion energy interaction equations have been modified to better describe the misfit and energy interaction terms between ions and solvent molecules. In addition, the salt solubility database is extended with additional non-aqueous systems containing solvents that have a low ($\epsilon_s < 15$) dielectric constant and promote near to full ion association. Throughout this work it is demonstrated that liquid-liquid equilibrium calculations and solid-liquid equilibrium predictions for electrolyte systems can be markedly improved with the inclusion of Bjerrum treatment based phenomenological considerations while introducing only one general additional parameter. Our modified approach reinforces the capabilities of COSMO-RS-ES as a powerful predictive tool for the calculation of phase equilibria in systems with scarce experimental data.

*Keywords: COSMO-RS, COSMO-RS-ES, electrolyte systems, liquid-liquid equilibria, solid-liquid equilibria, phase equilibria, modelling, low permittivity, Bjerrum association*

## 1. Introduction

Salts can have a large impact on the thermodynamics of chemical systems. For this reason, in some cases they create challenges for chemical engineering solutions when these are not considered in the development phase of a process. In other circumstances, they provide the very means to influence phase equilibria for a process to be viable in the first place. For both cases, reliable thermodynamic models are needed to accurately calculate a salt's influence on a given system.

The modelling of electrolyte systems started with $g^E$-model based approaches focused on aqueous systems mainly at low concentrations. More recent $g^E$-models with a wider applicability range are usually based on group contributions such as LIFAC/LIQUAC[1] and AIOMFAC[2] or on models that use binary interaction parameters such as e-NRTL[3]. Traditionally, $g^E$-models are a sum of at least two contributions. One describes the ion-ion interactions over longer distances (long-range term) and the other describes the direct interactions between the species (short-range term). In most cases full dissociation is assumed for the calculation of the long-range term (either Debye-Hückel or Mean Spherical Approximation based). Since these models are only valid at low concentrations, to be able to apply the model at higher salt concentrations, the specific ion effects are taken into account by one of the other terms[4–6]. The success of this approach in chemical engineering models spans over decades, being the Pitzer equations an influential example as these are sometimes included in other electrolyte $g^E$-models as an additional term to enhance their performance[1,7]. For this purpose, usually system-specific or binary interaction parameters are introduced. This increases the flexibility and adaptability of the models to adequately fit to a wide range of types of experimental data but reduces their predictive power.

Equations of state (EOS) such as eCPA[8–10], ePC-SAFT[11,12] and various SAFT-VR versions[13–15] have also been applied successfully to the calculation of electrolyte systems. The EOS can even be used to predict the phase behavior while avoiding the use of binary interaction parameters, but this always goes along with a decrease in accuracy. Modelling strategies with equations of state (EOS) also follow an approach where the Helmholz free energy sum includes several terms. To the extent of our knowledge, most of the electrolyte EOS employ a Debye-Hückel or MSA based long-range term analogous to the ones used in the $g^E$-models. However some may[11,16] or may not[10,17] explicitly include ion-ion dispersion interactions to describe the short-range effects. An exemplary case is the evolution of the PC-SAFT equation of state for electrolytes (ePC-SAFT[18]), which was applied by Held et al[19] with an implementation that considered ion pairing in weak electrolytes and treated the association constants and the ion pair segment diameters as additional parameters. The revised ePC-SAFT model[20] proposed a new modelling approach which explicitly included cation-anion dispersion interactions and was able to describe a wide variety of different electrolyte systems.

In spite of these successful strategies, explicit ion pairing is commonly neglected in thermodynamic modelling and accounting for its effects usually relies on other type of adjustments, thus hindering the capability for extrapolations and predictions especially for systems where it becomes prominent[21].

For neutral systems, a model that is able to calculate phase equilibria without binary interaction parameters is COSMO-RS[22,23]. In our previous publications[4,24], we have shown that the COSMO-RS based predictive electrolyte $g^E$-model COSMO-RS-ES is capable of describing a wide variety of mixed solvent systems. These include mean ionic activity coefficients (MIAC), liquid-liquid equilibria (LLE), Gibbs free energies of transfer of ions and solid-liquid equilibria (SLE). For most systems, the calculations of the model agree very well with experimental data. However, especially for LLE and SLE systems in solvents with a low bulk permittivity and increasing salt concentrations, larger deviations from the experimental values have been observed. A possible explanation for these deviations is the existence of partial dissociation (ion pairing), given that this phenomenon is favored by higher concentrations and lower solvent static permittivity values.

Analogous to other electrolyte $g^E$-models, so far the COSMO-RS-ES model assumes full dissociation for the long-range term. Due to its predictive character its flexibility to compensate for the deviations of the Debye-Hückel theory at higher concentrations is restricted. The present work demonstrates that the introduction of Bjerrum treatment based considerations for ion pairs into the long-range term leads to an overall improvement of the predictive capacity of COSMO-RS-ES.

## 2. Theory

## 2.1. **COSMO-RS-ES**

COSMO-RS-ES is built around a modification of the COSMO-RS model. It is based on the assumption that the short-range (SR) interactions between the species can be described by the COSMO-RS theory while the long-range (LR) interactions can be described with the Pitzer-Debye-Hückel (PDH) model as has been used in previous publications[4,24]. These terms are added according to the following equation:

$$\ln(\gamma_i) = \ln\left(\gamma_i^{SR,COSMO-RS}\right) + \ln(\gamma_i^{LR,PDH}) \quad (1)$$

COSMO-RS (COnductor like Screening MOdel for Realistic Solvation) uses quantum chemical information coupled with statistical thermodynamics to calculate the non-ideal interactions in mixtures.

PDH is a primitive model in the sense that it is based on the Debye-Hückel theory and describes the interaction of the ions as point charges interacting in a dielectric continuum characterized by its dielectric constant. For more information on the underlying theory of both models the reader is referred to the original publications[22,23,25,26].

## 2.2. **Ion pairing**

Ion pairing is commonly referred to as an explanation for the deviations observed in thermodynamic models that properly reproduce experimental data at low ionic strengths but lose their validity at high ionic strengths. The most widely accepted theory proposes the existence of solvent separated, solvent shared and contact ion pairs and that these behave as quasi-dipolar neutral species that do not contribute to the ionic strength of the solution and can be considered as nonelectrolytes that are being salted out by the free ions[27]. Evidence of the existence and influence of these species range from conductance studies[28,29] to EXAFS measurements[21]. It is also suggested in the literature that ion complexation-like phenomena at higher concentrations and low static permittivity values may take place to form triplets and quadrupoles[30], however the present work focuses mainly on a primitive model for symmetrical ion pairs.

For a 1:1 electrolyte the free ions and the ion pairs may be considered in equilibrium, as shown in Figure 1, where the right hand side of the equation is representative of the three types of ion pairs and $\mathrm{K}_A$ may be considered as an overall association constant.

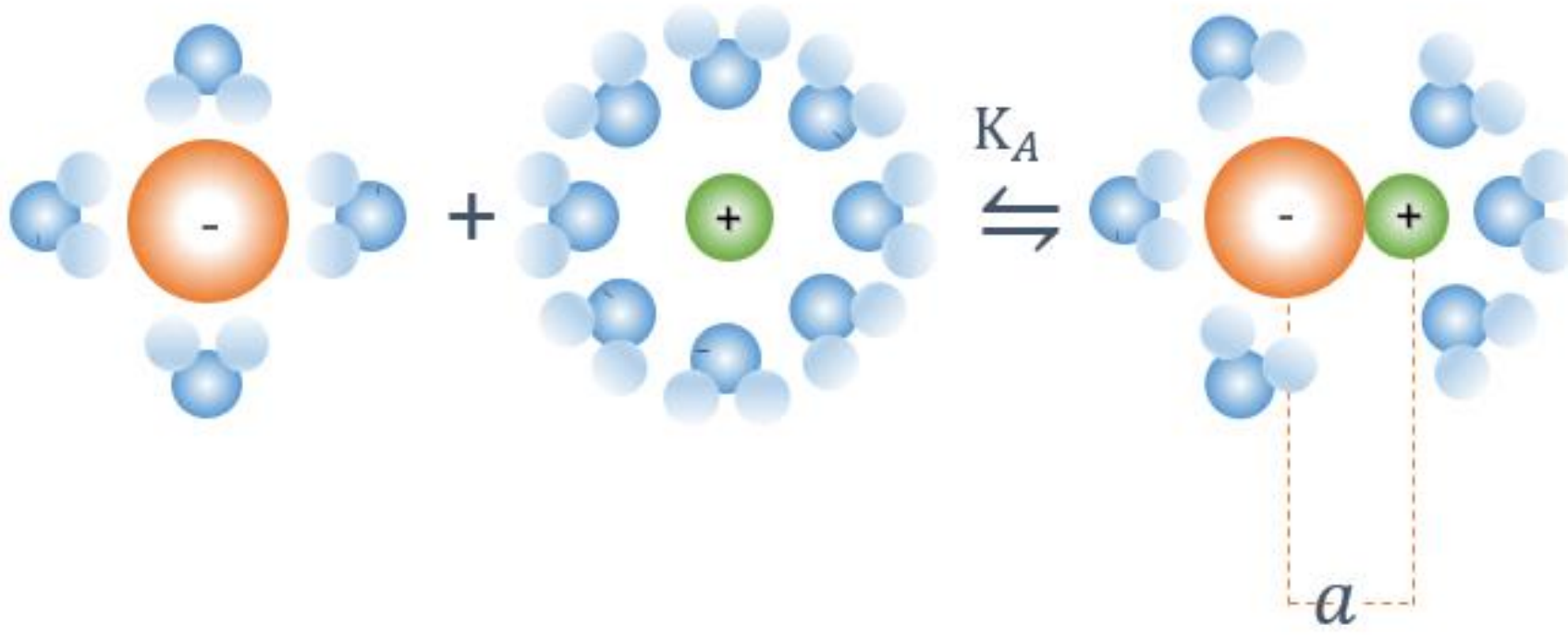

**Figure 1. Free ions and ion pairs in equilibrium.**

The equilibrium between the free and paired ions, assuming a single step reaction, can be expressed by the law of mass action as follows:

$$\mathrm{K}_A^{(\mathrm{x})} = \frac{(1-\alpha)\cdot\gamma_{IP}^{(x)}}{\alpha^{v}\cdot x_{\pm}^{v-1}\cdot(\gamma_{\pm,f}^{(x)})^{v}} \tag{2}$$

As an initial approach, the activity coefficient of the ion pair $\gamma_{IP}$ can be set to unity[30], although in recent literature the explicit value has been considered[11,19,31]. The term $v$ is the sum of the stoichiometric coefficients and the activity coefficient of the free ions $\gamma_{\pm,f}^{(x)}$ is naturally taken as a function of the concentration of the free ions[30]. When the dissociation degree is known (e.g. from conductivity studies) then the association constant may be calculated. In the absence of experimental data the dissociation degree may be estimated when some expression for the association constant is available[30,32].

For instance, the Coulombic association constant may be calculated through relations like those of Fuoss[33], Bjerrum[34] or Ebeling[35]. The Danish chemist Niels Bjerrum was the first to propose an equation based on the restricted primitive model to evaluate the association constant of ions embedded in a dielectric medium. Bjerrum found a minimum for the Boltzmann distribution based probability of the volume element of an ion encountering that of its counter-ion at a distance where the Coulombic interaction energy doubles the thermal energy $\mathrm{k_B}T$. This length was therefore considered as a cut-off criterion for the separation distance below which two ions are considered to be paired[34]. This distance is known as the Bjerrum length:

$$l_B = \frac{|z_i z_j| e^2}{4\pi\epsilon_0\epsilon_s(2k_BT)} \quad (3)$$

where $z$ is the valence of the ions, $e$ is the elementary charge, $\epsilon_0$ is the permittivity of vacuum and $\epsilon_s$ Is the solvent's dielectric constant. The Bjerrum length and a distance of closest approach (hereafter $a$) are the radii that define a volume element of the dielectric that surrounds an ion. If the volume elements of an ion and its counter-ion are close enough to come into contact, then from that point on the ions are considered to be paired regardless of whether they remain at such a distance or come closer together. Consequently, through geometrical and theoretical considerations[28,34] the following expression can be derived:

$$\mathrm{K}_A^{(c)}[\mathrm{M}^{-1}] = 4\pi N_A \cdot 1000 \cdot \int_a^{l_B} \exp\left(\frac{2l_B}{r}\right) r^2 dr \quad (4)$$

where Avogadro´s constant converts number density to molar concentration and the factor of 1000 arises from a conversion to liters if $a$ and $l_B$ are expressed in meters (equation (4) and its variants are commonly[27,28,30,33] written for lengths in centimeters).

For symmetrical electrolytes the association constant of a single step association/dissociation reaction may be converted to the mole fraction scale with the following expression:

$$\mathrm{K}_A^{(x)} = \mathrm{K}_A^{(c)} \cdot \left(\frac{\rho}{M}\right)^{v} \tag{5}$$

where $\rho$ and $M$ are taken as the density and molar mass of the mixture, respectively. In dilute cases they may be taken as pure solvent properties.[36]

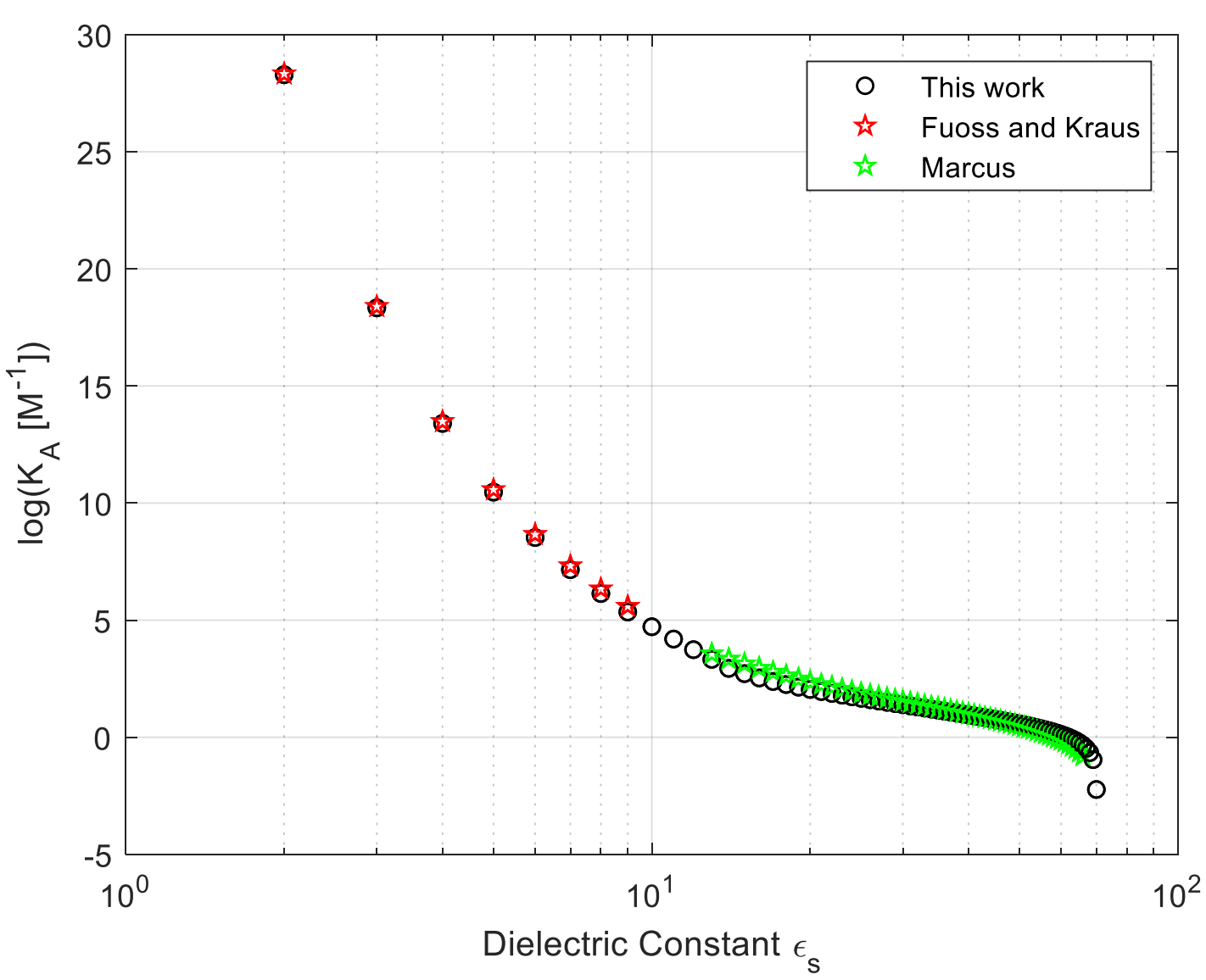

**Figure 2. $\mathrm{K}_A^{(c)}$ as a function of a solvent´s dielectric constant calculated for $a = 4$ Å with three polynomial approximations. Black circles: used in this work and valid for the whole range of $2l_B/a$; a polynomial[28] valid when $2l_B/a \leq 10$ and a polynomial[27] valid when $2l_B/a \geq 15$**

Figure 2 shows an association constant for $a = 4$ Å as a function of the dielectric constant of the solvent. It is evident that the association constant increases as the bulk permittivity decreases. Not accounting for these considerations could lead, for instance, to the underestimation of the concentration of a salt in an organic phase. A possible explanation for this is that ion pairs as dipoles play a role in raising the affinity of a salt towards a liquid phase[10,21].

### 2.3. **Calculation of phase equilibria**

For two liquid phases dividing into an organic phase $O$ and a salt-rich phase $S$, the following equation holds true:

$$x_i^O \gamma_i^O = x_i^S \gamma_i^S \tag{6}$$

Where $x_i$ is the mole fraction and $\gamma_i$ the activity coefficient of component $i$ in the respective phase. In the case of an electrolyte the MIAC would be employed.

By introducing the partition ratio $K_i^{OS}$ it might be reformulated like so:

$$K_i^{OS} = \frac{x_i^O}{x_i^S} = \frac{\gamma_i^S}{\gamma_i^O} \tag{7}$$

For a strong electrolyte, the solubility equilibrium condition might be formulated like so:

$$\ln K_{SP} = (\nu_{cat} + \nu_{an}) \ln(x_{\pm} \gamma_{\pm}) \tag{8}$$

Where $K_{SP}$ is the solubility product, $x_{\pm}$ is the mean ionic mole fraction and $\gamma_{\pm}$ is the mean ionic activity coefficient. In this work, the solubility product is predicted with equation (8) using the experimental solubility in water and the activity coefficient in water calculated by the model.

The Gibbs free energy of transfer is calculated as the ratio of the ion activity coefficients at infinite dilution in water (W) and in the organic solvent (O):

$$\Delta G_{ion}^{\mathrm{t},W \to O} = RT \ln \left( \frac{\gamma_{ion}^{\infty,\mathrm{O}}}{\gamma_{ion}^{\infty,\mathrm{W}}} \right) \tag{9}$$

## 3. Computational details of COSMO-RS-ES

### Short-range model: modified COSMO-RS-ES

The neutral species are described with the misfit term using the misfit correlation and the hydrogen bonding energy term. For the contact between ions or between an ion and a solvent the equations from Table 1 are employed. These equations are a modified version of the original

COSMO-RS-ES where sigma thresholds have been applied for most energy interactions and the sigma thresholds for polyatomic anions and anion-water interactions were dropped. In addition, the hydrogen bonding sigma threshold was applied for water. For further details regarding our implementations of COSMO-RS based electrolyte models please refer to the first publication of COSMO-RS-ES[4].

**Table 1: Modified interaction energy equations for ionic interactions in the short-range contribution of COSMO-RS-ES.**

| Interaction | Misfit Factor | Ionic Interaction Energy Term |
|---|---|---|
| cation – $H_2O$ | $A_1$ | $E^{ion}_{cat-H_2O} = \frac{a_{eff}}{2} B_1 \sigma_{cat} \max\left(0, \sigma_{H_2O} - \sigma_{HB}\right)$ |
| cation – org. mol. | $A_2$ | $E^{ion}_{cat-om} = \frac{a_{eff}}{2} B_2 \sigma_{cat} \max(0, \sigma_{om})$ |
| cation - halide | 0 | $E^{ion}_{cat-hal} = \frac{a_{eff}}{2} B_3 \min\left(0, \sigma_{cat}(1 - D_1 \lvert\sigma_{cat}\rvert^{E_1})\right) \sigma_{hal}$ |
| cation – polyat. an. | 0 | $E^{ion}_{cat-pa} = \frac{a_{eff}}{2} B_4 \min\left(0, \sigma_{cat}(1 - D_1 \lvert\sigma_{cat}\rvert^{E_1})\right) \max\left(0, \sigma_{pa}\right)^{E_2}$ |
| halide - $H_2O$ | $A_3$ | $E^{ion}_{hal-H_2O} = \frac{a_{eff}}{2} B_5 \min\left(0, \sigma_{H_2O} + \sigma_{HB}\right) \max(0, \sigma_{hal})$ |
| halide – org. mol. | $A_4$ | $E^{ion}_{hal-om} = \frac{a_{eff}}{2} B_6 \min(0, \sigma_{om} + C_1) \max(0, \sigma_{hal} - C_2)^{E_3}$ |
| polyat. an. - $H_2O$ | $A_5$ | $E^{ion}_{pa-H_2O} = \frac{a_{eff}}{2} B_7 \min\left(0, \sigma_{H_2O} + \sigma_{HB}\right) \max\left(0, \sigma_{pa}\right)$ |
| polyat. an. - org. mol. | $A_6$ | $E^{ion}_{pa-om} = \frac{a_{eff}}{2} B_8 \min(0, \sigma_{om} + C_1) \max\left(0, \sigma_{pa}\right)$ |

**Long-range model: Pitzer-Debye-Hückel + Bjerrum Ion Pairs**

A volume fraction mixing rule for the permittivity and a salt-free mole fraction based mixing rule for the density neglecting excess volume effects in both cases were used when the system contained more than one solvent. For the closest approach parameter a value of 14.9[25] is fixed.

Commonly in the literature, when full dissociation is assumed, the activity coefficient is corrected based on the assumption $\gamma_{\pm} = \alpha \cdot \gamma_{\pm,f}$ in MSA or DH based approaches that consider association[35,37,38]. In the case of COSMO-RS-ES the short-range term considers ion-ion energy interactions, as shown in Table 1, thus inherently accounting for a distinction between the effects of ion pairs and free ions. Because COSMO-RS-ES inherently considers direct contact interactions, the ion association effects are only considered for the long-range term.

The correction strategy for ion pairing was implemented for the long-range term by retaking the concept of an "effective" ionic strength as function of ion concentration $c$ and charge $z$ which proposes the following[39]:

$$I_e = \frac{1}{2}\left\{\sum_f c_f z_f^2 + \sum_p c_p z_p^2\right\} \tag{10}$$

where $f$ and $p$ denote free and charged paired ions, respectively. For the case of symmetrical electrolytes with neutral ion pairs ($z_p = 0$) the effective ionic strength is given by $I_e = I \cdot \alpha$, where $I$ is the ionic strength assuming full dissociation. This effective ionic strength is then applied to the Pitzer-Debye-Hückel term as an adjustment.

For the purpose of estimating the dissociation degree $\alpha$, equations (1), (2), (4) and (5) can be applied iteratively until the law of mass action converges. Equation (1) was applied in each step by recalculating the concentration of the free ions with the iterated values of $\alpha$ and a threshold of $|\alpha_i - \alpha_{i-1}| \leq 0.005$ was selected as convergence criterion for each data point.

To keep the calculations as computationally inexpensive as possible, all salts were treated equally and only one association step was considered for asymmetric electrolytes, i.e. charged pairs and clusters are, for the moment, not being considered.

Finally, in our present approach, to keep the model simple and not go into the complexities of calculating the activity coefficient of the ion pairs, as in other publications[30,31], the value was set equal to unity.

**Parameterization and evaluation procedure**

The database used to parametrize the model includes MIAC systems (1190 data points) and LLE systems (1126 data points) at 25 °C and Gibbs free energies of transfer of ions from water to organic solvents (992 data points). An overview of the data used for training and evaluating the model is included in the supporting information. The initial point for the parameterization presented in this work are the values from our previous work[24] and a value of 0.0085 e/Å$^2$ for solvent sigma threshold values.

A Levenberg-Marquardt algorithm was employed to optimize all parameters. These include the radii of the cations, the parameters from Table 1 and a scaling factor $f_{scale}$ used for the calculation of the closest approach distance $a$ from equation (4), which is calculated as the sum of the ionic radii from COSMO-RS-ES multiplied by the scaling factor. On average, the ionic radii from COSMO-RS-ES and the ionic radii in solution reported by Marcus[40] deviate by a factor of 0.76. This value was applied as an initial guess for $f_{scale}$.

The following contribution to the objective function for MIAC systems was used in this work:

$$F_{MIAC} = W_{MIAC} \sum_{l} \left( ln\, \gamma_{\pm}^{(m),calc} - ln\, \gamma_{\pm}^{(m),exp} \right)^2 \qquad (11)$$

The weighting factor $W_{MIAC}$ varied depending on the system type and were taken as the same as the previous COSMO-RS-ES publications[4,24]. For alkali halides a value of 40, for all other alkali salts a value of 20 and for the alkaline earth salts a value of 5 was used. With the corresponding AAD:

$$AAD_{MIAC} = \frac{1}{N_{DP}} \sum_{i} \left| ln\left(\gamma_{\pm}^{(m),calc}\right) - ln\left(\gamma_{\pm}^{(m),exp}\right) \right| \qquad (12)$$

To allow an efficient evaluation of the models´ capabilities to predict the solubility of salts in this work, the expected MIAC that the model should deliver to calculate the correct solubility is compared to the value actually calculated.

Consequently, the deviation for the SLE systems was calculated as follows:

$$AAD_{SLE} = \frac{1}{N_{DP}} \sum_{i} \left| ln\left(\gamma_{\pm,i,other}^{+,expected}\right) - ln\left(\gamma_{\pm,i,other}^{+,calc}\right) \right| \tag{13}$$

with

$$ln\left(\gamma_{\pm,i,other}^{+,expected}\right) = \ln\left(x_{\pm,ref}^{exp} \cdot \gamma_{\pm,ref}^{+,calc}\right) - \ln(x_{\pm,i,other}^{exp}) \tag{14}$$

where water is taken as the reference solvent for all salts.

The deviation for the LLE systems was calculated based on partition coefficients of the salt between the salt-rich and organic phase as follows:

$$AAD_{LLE} = \frac{1}{N_{DP}} \sum_{i} \left| ln\left(K_i^{OS,calc}\right) - ln\left(K_i^{OS,exp}\right) \right| \tag{15}$$

where the salt-rich phase is denoted by $S$ and the organic phase by $O$.

For Gibbs free energies of transfer, the overall deviation is given by:

$$AAD_{G-transfer} = \frac{1}{N_{DP}} \sum_{i} \left| \frac{\Delta G_{ion}^{\mathrm{t},W\rightarrow O}}{RT} - \ln\left(\frac{\gamma_{\mathrm{ion}}^{\infty,\mathrm{O}}}{\gamma_{\mathrm{ion}}^{\infty,\mathrm{W}}}\right) \right| \tag{16}$$

In this work whenever the data was used in the parameterization process, the results are considered to be correlations and whenever the data shown was not part of the training set the results are considered to be predictions. The results presented in the first COSMO-RS-ES publication[4] are taken to compare LLE correlations and the results presented in the second COSMO-RS-ES publication[24] are taken to compare SLE predictions.

## 4. Results

### 4.1. Importance of Ion pairing

Figure 3 shows the deviation of the calculated SLE systems, given by $ln\left(\gamma_{\pm,i}^{expected}\right)$ - $ln\left(\gamma_{\pm,i}^{calc}\right)$, using the last published version of COSMO-RS-ES[24] (taken for comparison). It can be observed that the systematic underestimation correlates with high salt concentrations and low solvent dielectric constant values.

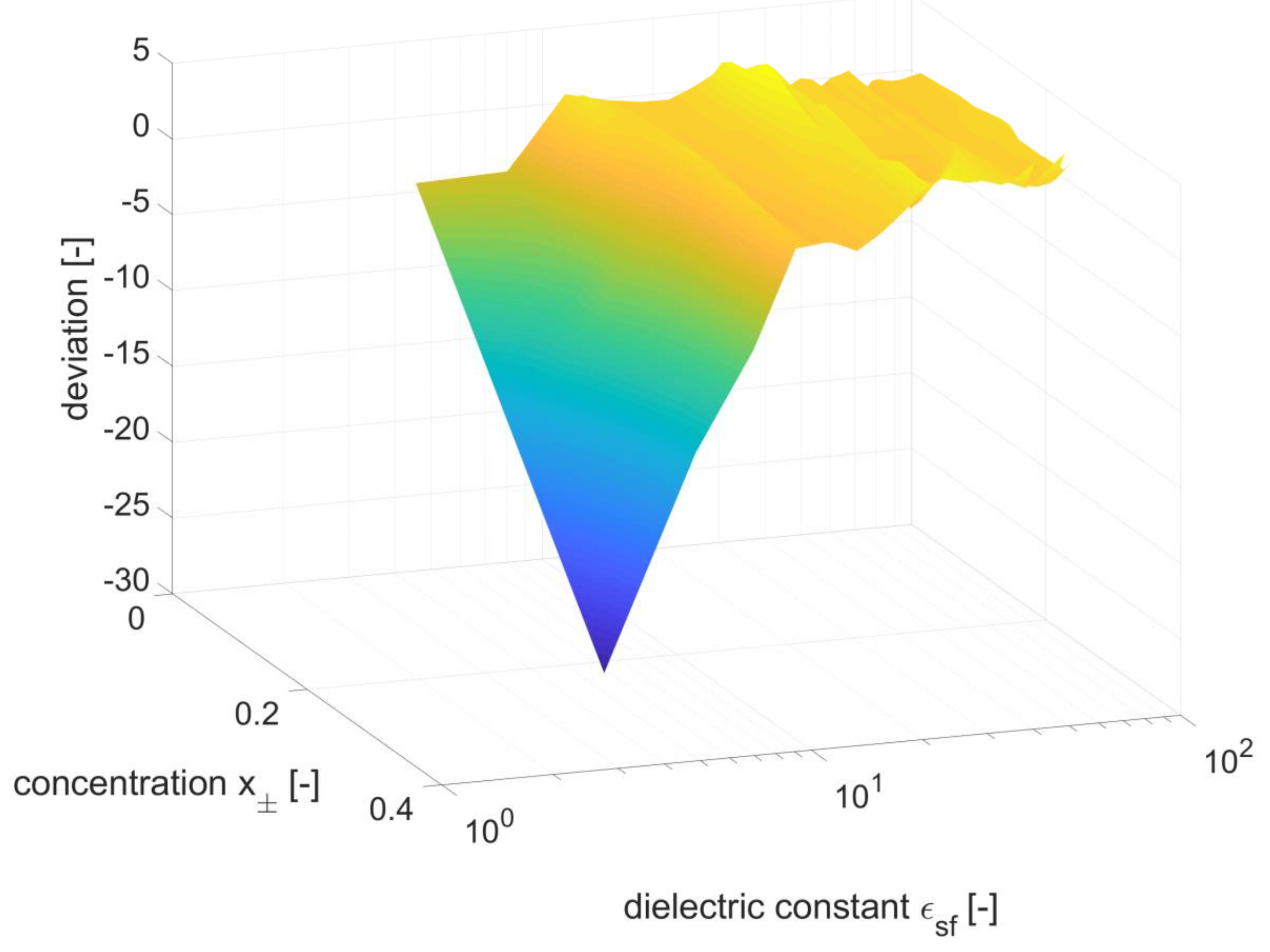

**Figure 3. Residual deviation of SLE systems as a function of solvent dielectric constant and salt concentration.**

These observations also hold for LLE systems where large underestimations are found for mixtures containing highly soluble salts like lithium chloride in solvents with a very low static permittivity (i.e. 2-methyl-2-butanol). Since these underestimations do not exist for the case of calculations performed for Gibbs free energies of transfer of ions at infinite dilution, it becomes clear, that the deviations must originate due to an effect at higher salt concentrations.

The previous observations serve as a hint that the systematic underestimation of LLE and SLE systems is more dependent on the long-range term than the short-range term. Given that salts tend to associate with rising salt concentration and decreasing static permittivity of the medium[30], an adjustment to the long-range term presents itself as a possible improvement to further refine the model.

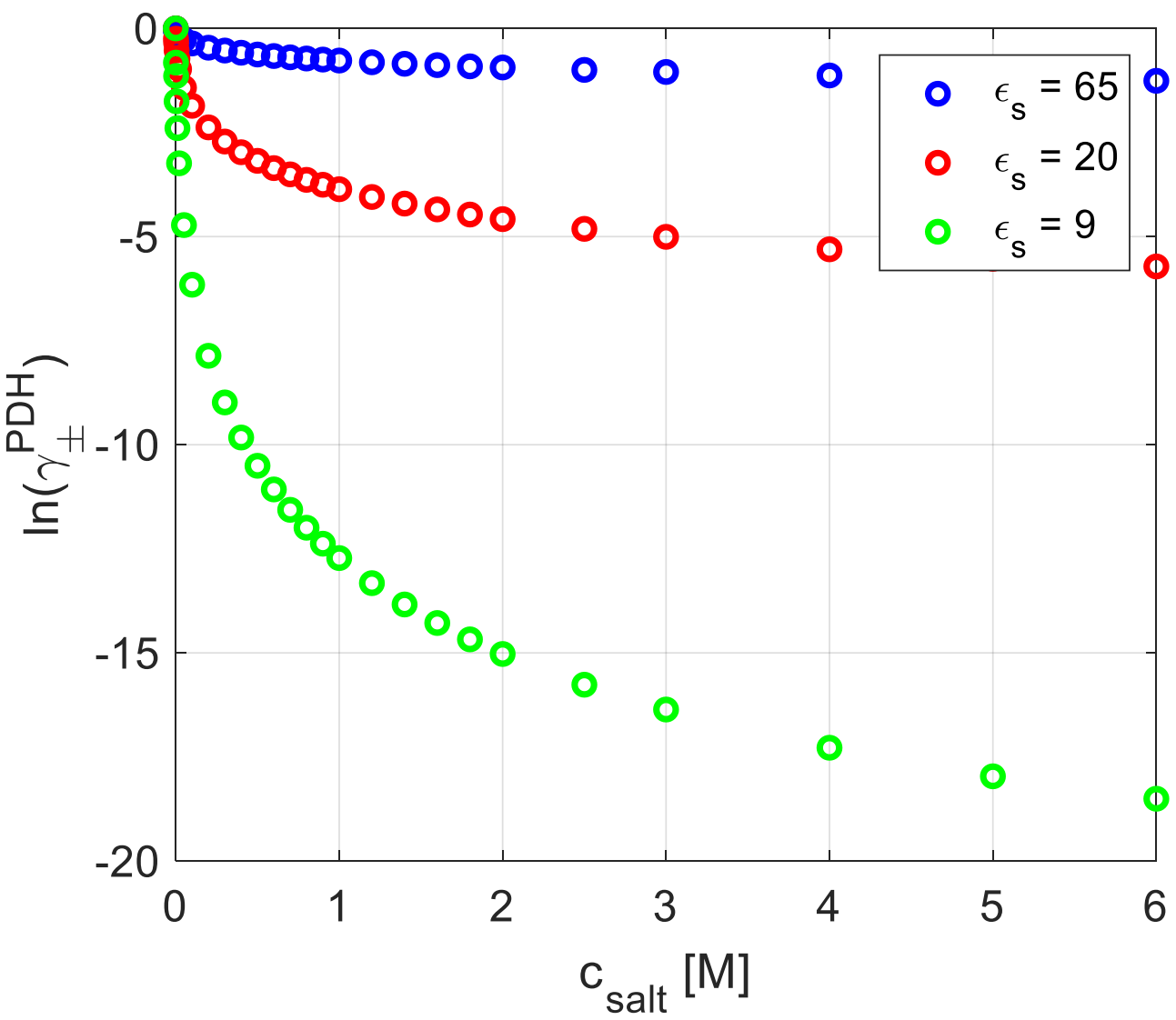

**Figure 4. Behavior of the Pitzer-Debye-Hückel term as function of solvent permittivity for a hypothetical fully dissociated 1:1 electrolyte.**

Evaluating the effect of the solvent´s static permittivity on the PDH term is helpful in understanding how this widespread electrolyte model works in solvents with low static permittivity. While it is known that the Pitzer-Debye-Hückel derivation has theoretical limitations like considering the permittivity and density as salt concentration independent, these limitations are accounted for by the parameterization of the other terms. For instance, it may be appreciated in Figure 4 that the full dissociation assumption leads to a dramatic decrease of the logarithmic activity coefficient as static permittivity values become lower and it may be safely stated that full dissociation of an inorganic salt in a medium with $\epsilon_s = 9$ is not a reasonable assumption. Many models like the Pitzer equations or those with Pitzer-like terms have the adaptability to counter-balance the deviations arising due to these effects. The case is different for a COSMO-RS based approach where the underlying principles and phenomenology of electrolyte solutions like extensive ion pairing[30] or the behavior of the decay length in concentrated electrolytes[41,42] cannot be easily fitted while at the same time keeping the model as universal and predictive as possible. COSMO-RS-ES focuses more on the correct reproduction of global trends than on very high precision system-specific fitting. Therefore, the underlying phenomena must be revisited if higher predictive quality is desired without the need for system or chemical group specific parameters.

### 4.2. Refined Model including Ion pairing

Introducing considerations for ion pairing into the model increases the model complexity and the computational effort since the calculation must be done iteratively. Due to the high sensitivity of the model towards the scaling factor $f_{scale}$, several strategies were considered for the subsequent parameterization. It was found that the most effective parameterization is done in several steps: first a fixed initial scaling factor of 0.76 is set to estimate the distance of closest approach in Bjerrum's equation and an optimization is performed for the parameters in Table 1. Subsequently, the new optimized parameters from Table 1 are fixed and the radii and scaling factor used to calculate the association constant are then optimized. Finally, a refinement step for all the parameters from Table 1 is performed with the new radii and scaling factor from the second step. The resulting parameters of this parameterization are shown in Table 2.

**Table 2. Parameters for COSMO-RS-ES including an ion pair treatment for the long-range term.**

| cation radii [Å] scaling factor[−] | | Parameters A,B $\left[\frac{\text{kJ Å}^2}{\text{mol e}^2}\right]$ | | parameters C $\left[\frac{\text{e}}{\text{Å}^2}\right]$ | | parameters D,E [−] | |
|---|---|---|---|---|---|---|---|
| $Li^+$ | 1.697 | $A_1$ | 5515 | $C_1$ | 0.0096 | $D_1$ | 1852 |
| $Na^+$ | 1.874 | $A_2$ | 4151 | $C_2$ | 0.0097 | $E_1$ | 2.075 |
| $K^+$ | 1.993 | $A_3$ | 3965 | | | $E_2$ | 5E-06 |
| $Rb^+$ | 2.071 | $A_4$ | 3294 | | | $E_3$ | 0.036 |
| $Cs^+$ | 2.231 | $A_5$ | 3404 | | | | |
| | | $A_6$ | 3802 | | | | |

| | | | |
|---|---|---|---|
| $f_{scale}$ | 0.824 | $B_1$ | 13554 |
| | | $B_2$ | 166 |
| | | $B_3$ | 3795 |
| | | $B_4$ | 30 |
| | | $B_5$ | 20497 |
| | | $B_6$ | 789 |
| | | $B_7$ | 12302 |
| | | $B_8$ | 13624 |

The parameterization presented here is capable of reproducing the trends of Gibbs free energies of transfer of ions with a similar accuracy compared to the parameterization from our previous work[24]. The trends lie within the experimental error and exhibit no qualitative differences.

The case is similar for MIAC values of aqueous solutions. The accurate evaluations for this type of systems remain practically unchanged. The value for calculations with the parameters from Müller et al.[24] is $AAD_{MIAC} = 0.0429$ while the value for the calculations with the parameters presented in this work is $AAD_{MIAC} = 0.0425$.

The deviation of the calculated LLE systems has been further reduced by 25% from a value of $AAD_{LLE} = 0.69$ to a value of $AAD_{LLE} = 0.52$ in this work. This improvement is partly driven by the introduction of threshold values for the organic solvent in several equations from Table 1 and by the adjustment of the ion pair treatment in systems with a very low static permittivity (i.e. 2-methyl-2-butanol, 3-pentanol and ethyl acetate, all of which have a dielectric constant below 15). These results can be visualized in the parity plots presented in Figure 5.

Systematic deviations had been observed in previous versions of the model for halide salts in acetonitrile containing systems. These are the systems that exhibit marked deviations in Figure 5 (left) corresponding to the LiCl, NaBr and KBr in water + acetonitrile reported by Renard and Heichelheim[43,44] and Renard and Oberg[45]. In these systems the organic phase has a low salt concentration (below 3 wt%) and acetonitrile is a polar aprotic solvent, therefore the correction for these systems is mainly due the modifications of the short-range term. As a supporting argument, dissociation degrees of 80% or higher were found for these systems, thus confirming that the correction is mostly given by the short-range term.

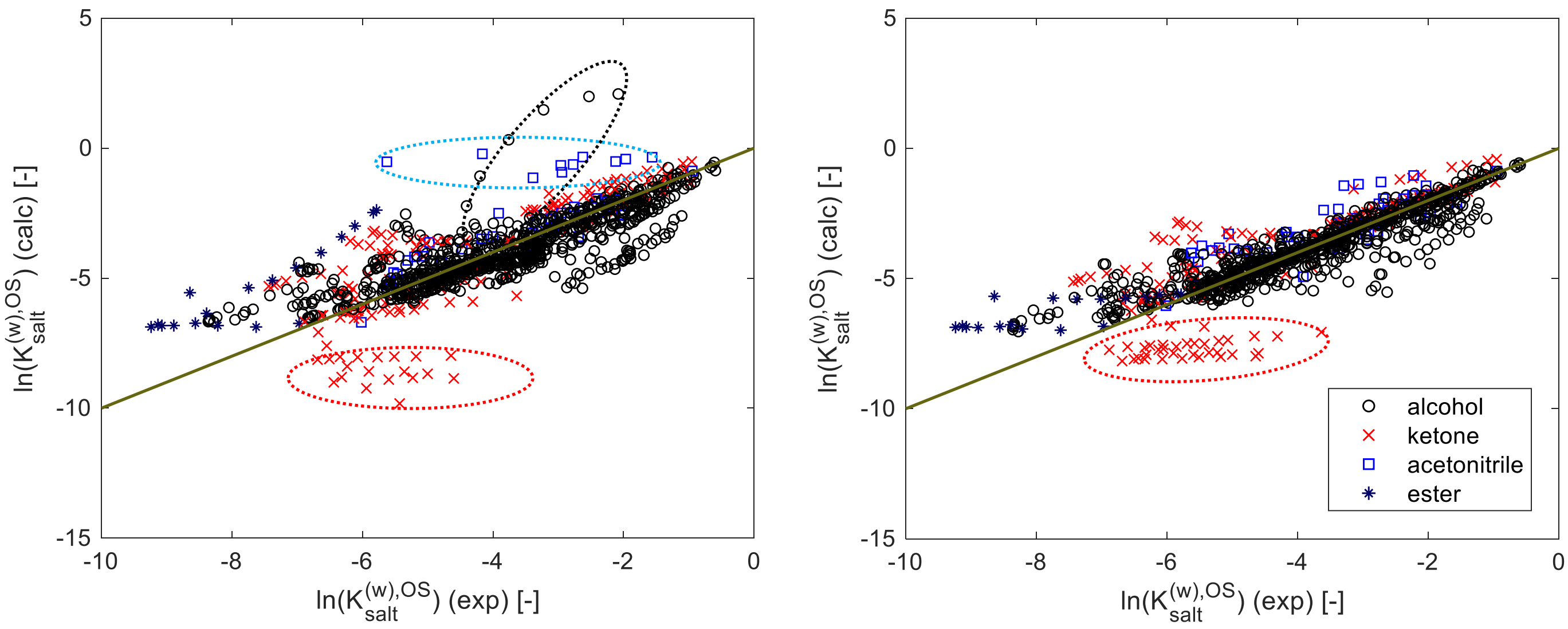

**Figure 5. Parity plots of the logarithmic partition coefficients of diverse salts in water / organic solvent mixtures. Left: calculated with the original COSMO-RS-ES[4]. Right: this work. Black oval: LiCl + water + 2-methyl-2-butanol system. Blue oval: water + acetonitrile systems (NaBr, KBr, LiCl). Red oval: systems containing MIBK or 2-butanone.**

In contrast to the aforementioned systems, the corrections for systems where the static permittivity of the organic phase is very low (e.g. 2-methyl-2-butanol $[\epsilon_s = 5.8]$ ) were mainly driven by a reduction of the long-range contribution which is now calculated with the effective ionic strength as a function of the dissociation degree.

Figure 6 shows the partition coefficients corresponding to the system $KNO_3$ + water + 2-methyl-2-butanol system as a function of salt concentration in the aqueous phase. An improvement in both the partition coefficients of the salt and the organic solvent may be seen in the results of this work. This is due to the fact that the chemical potentials of each of the species are connected via the Gibbs-Duhem relation, therefore an improvement that exerts an impact on the activity of one component has an indirect effect on the partition of all other components. Furthermore, this is a system where, even though the actual salt concentration in the organic phase is small (below 0.22 wt%), the considerably low static permittivity of the organic phase remains near that of the organic solvent and ion pairing would be expected. The algorithm converged at near full association ($\alpha^{O} \rightarrow 0$) for the organic phase whereas the salt remains dissociated in the salt-rich phase ($\alpha^{S} \rightarrow 1$).

Figure 6 also shows exemplarily how different versions of the COSMO-RS-ES model have evolved to better calculate this system. Introducing Gibbs free energies of transfer in the training set[24] improved the performance and the model has been made more accurate by the introduction of ion pairing considerations for the long-range term.

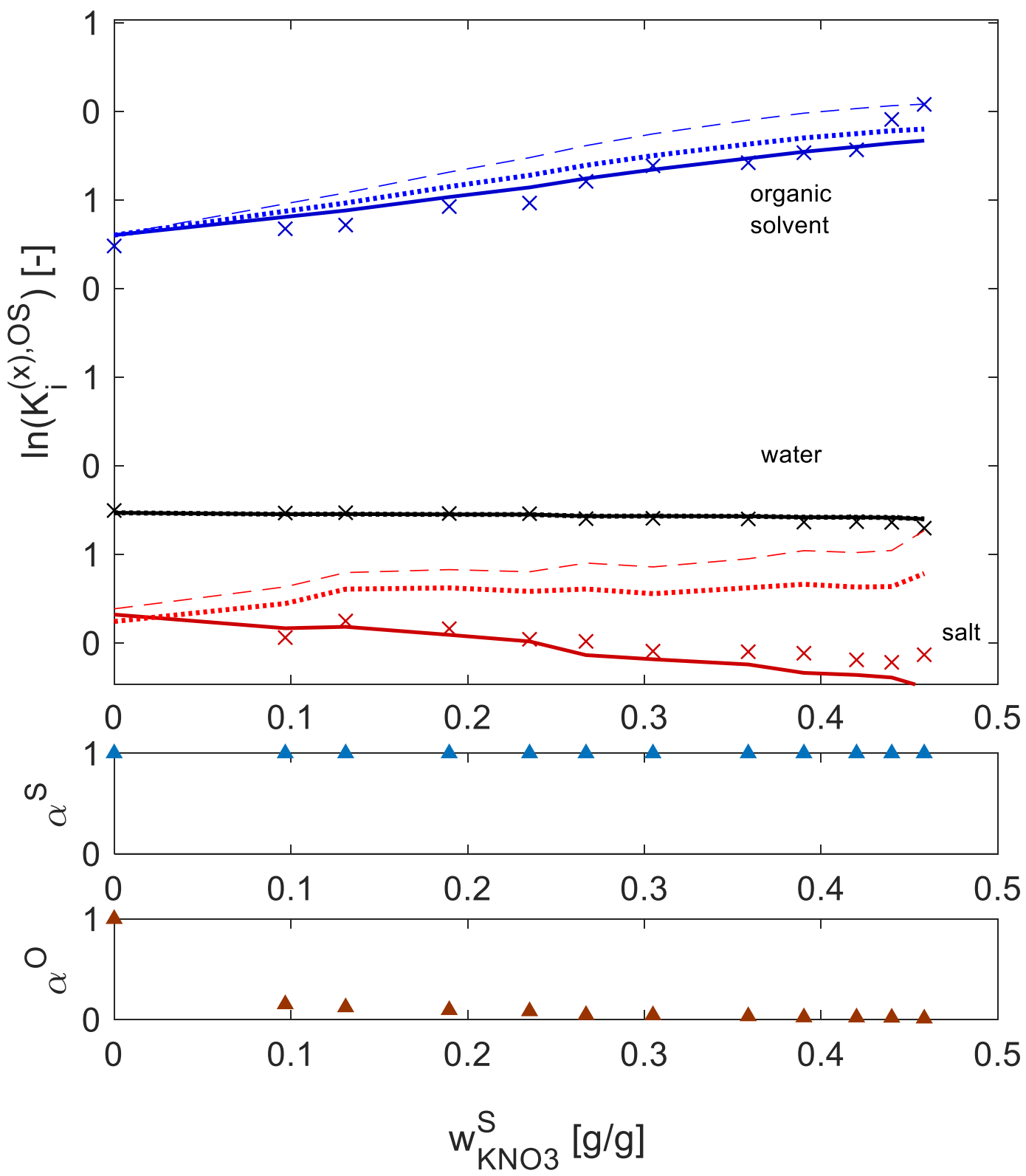

**Figure 6. Calculated and experimental (×) logarithmic partition coefficients for the $KNO_3$ + water + 2-methyl-2-butanol system. Dashed line: calculated with parameters from Gerlach et al[4]. Dotted line: calculated with parameters from Müller et al[24] . Solid line: this work. Subplots show the salt´s dissociation degree $\alpha$ in the organic ($O$) and salt-rich ($S$) phases.**

In spite of the iterative calculation for the law of mass action using equation (1), short-range non-coulombic contributions to the association constant are ignored by the present approach. From this perspective, it was the objective of this work to extend the model with relatively simple and effective considerations instead of more intensive or computationally expensive approaches. In other words, the ion pairing treatment and dissociation degrees presented here have to be interpreted under the light of them being an approach for better modelling performance and have to be treated carefully when analyzing the phenomenology of a given particular system. For instance, it may be observed in Figure 6 on the right side that the trends are reverted in the last two data points: the salt seems to prefer the organic phase past a certain point. Whether this can or cannot be attributed to experimental error lies beyond the scope of this work. However, it could be attributed to specific effects like the formation of ion clusters in the organic phase. Also ion pairing may increase the affinity of the salt for an organic phase[21,30]. Even though our model is not explicitly accounting for charge imbalances in ion aggregates, formation of triplets in a 1:1 electrolyte or formation of charged ion pairs in asymmetrical electrolytes, the present results show evident improvements.

Some small but systematic deviations can still be observed for systems with a polar aprotic solvent with a larger molecular weight than DMSO or acetonitrile e.g. alkali-halide salts in 2-butanone and methyl-isobutyl-ketone (MIBK) or salts in ethyl acetate. This might present an opportunity to reassess the COSMO-RS-ES equations in the future to further improve the description of these systems.

COSMO-RS-ES has so far only been adjusted to thermodynamic data of ions. This could possibly lead to training the model for this specific target property but failing to reproduce the thermodynamic properties of the other components in the different mixtures. In the case of the LLE systems studied in this work, this can be tested by looking at the predicted partition coefficients of the solvents compared to the experimental values. As can be seen in Figure 7, the overall prediction of the partitioning coefficients of the neutral species does not deteriorate but in most cases even improves further with the introduction of the developments reported in this work. This shows the thermodynamic consistency of the model as the description of all activity coefficients improves and not only the ones of the salt.

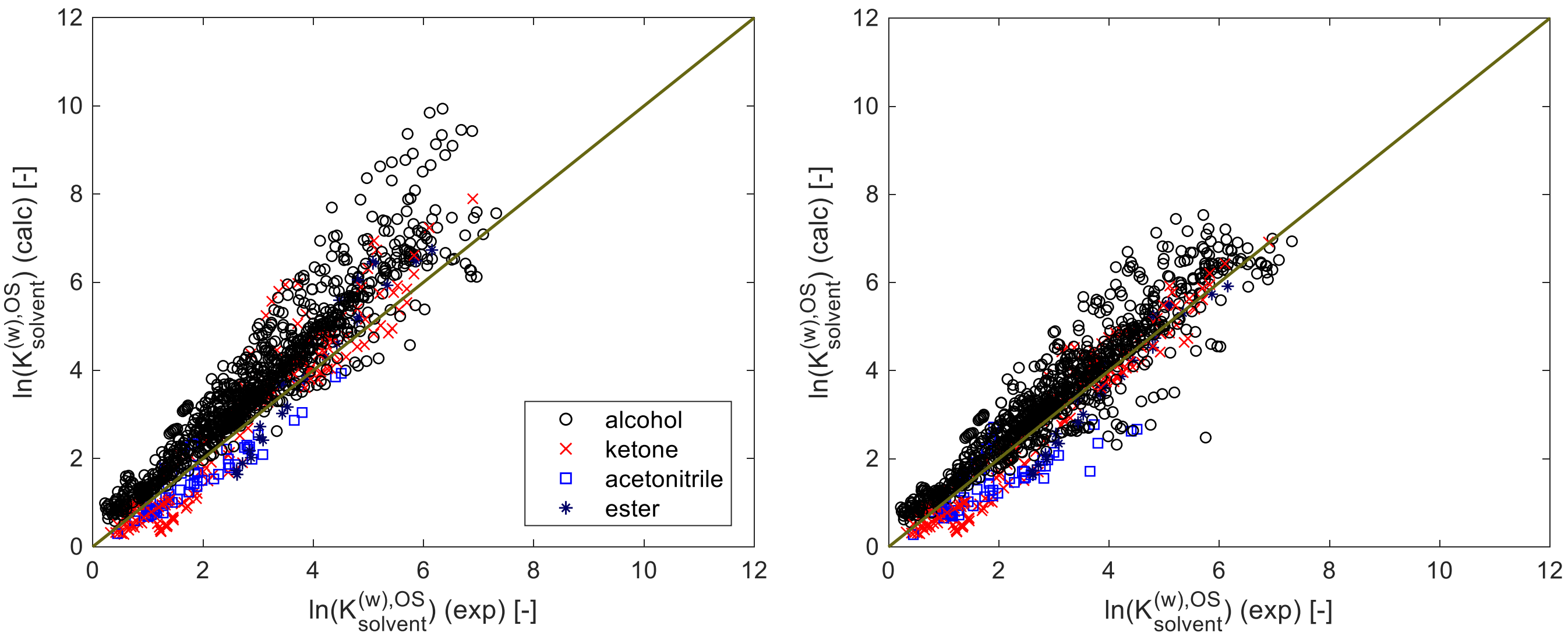

**Figure 7. Parity plots of the logarithmic partition coefficients of diverse organic solvents in salt + water / organic solvent interfaces. Left: calculated with the original COSMO-RS-ES[4]. Right: this work.**

Overall, the model has been improved for most LLE systems showing the importance of describing ion pairing already at moderate salt concentrations.

As an example for an LLE calculation, Figure 8 shows the salting-out of MIBK from the salt-rich phase as reported by Mohammad et al.[46] comparing the calculations of ePC-SAFT to the calculations of this work. COSMO-RS-ES is able to deliver a correct qualitative description of the trends and in some cases even a quantitative description of the system. It must be noted however, that the salt-free water/MIBK system already presents a clear overestimation with respect to the experimental partitioning of MIBK. This deviation is carried over to the calculation of the mixtures that include salts. This overestimation goes in line with the observations made for ketone containing systems in Figure 5 and Figure 7.

So far, in the development of the COSMO-RS-ES model, emphasis has been put exclusively into describing the interactions with ions more accurately. Perhaps a point has been reached, where it becomes relevant to describe the interactions of the neutral components in a better way to improve the overall performance of the model.

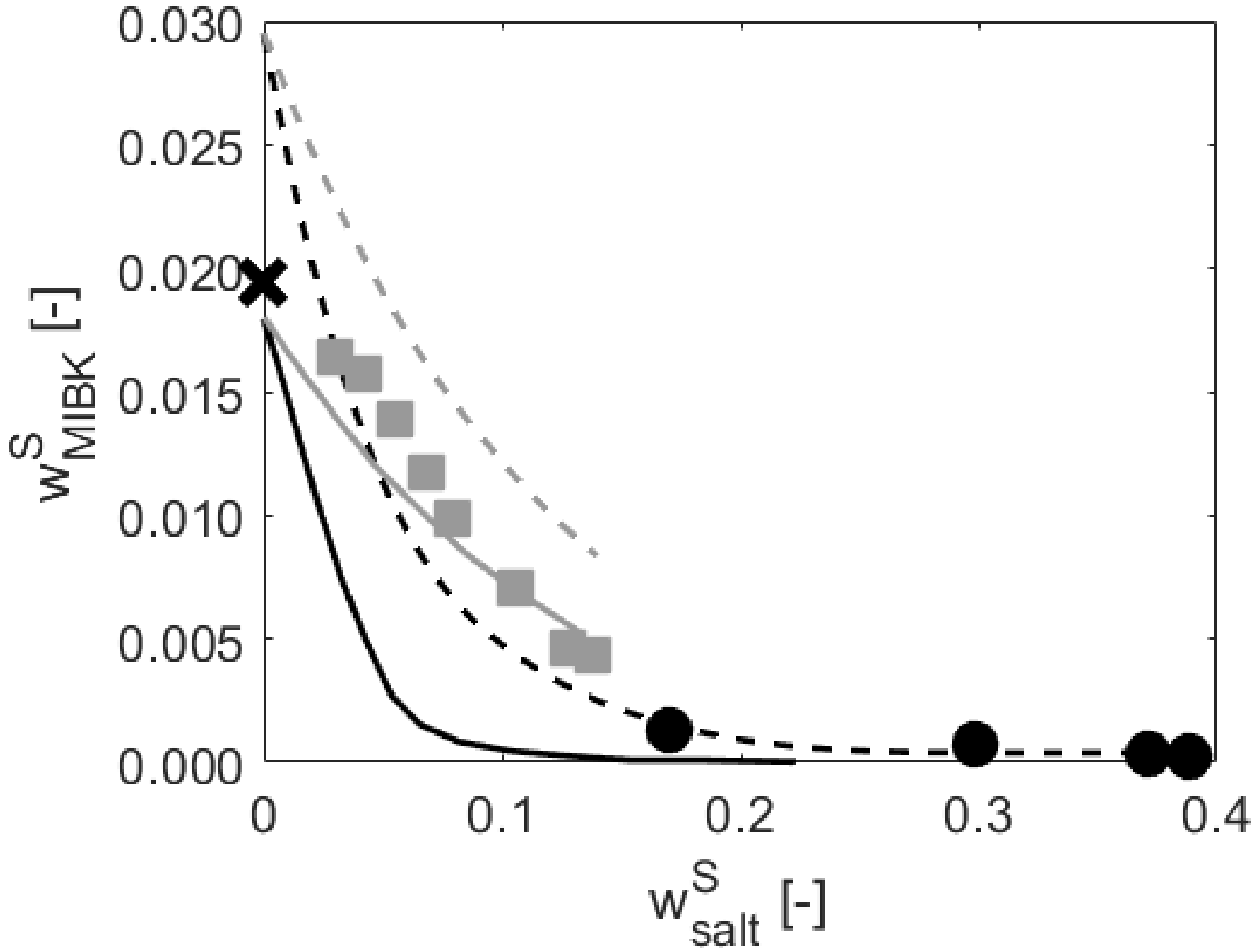

**Figure 8: Weight fraction of MIBK in the salt-rich phase vs. salt weight fraction in the salt-rich phase. Circles (LiCl + MIBK + water) and squares ($Na_2SO_4$ + MIBK + water) are experimental results from Mohammad et al.[46] while the cross (MIBK + water) is from Yang et al.[47] Full lines are modeling results with ePC-SAFT[46] and dashed lines are calculations with COSMO-RS-ES. Grey lines: $Na_2SO_4$, black lines: LiCl.**

### 4.3. Prediction of SLE Systems

The database of salt solubility data presented in our previous work[24] has a limited number of data points with solvents that have a low static permittivity. Some additional measurements with these types of solvents have been added to the database. The evaluated SLE database consists of 835 data points from which 43 contain a solvent with a dielectric constant of 15 or lower.

The deviation of the calculations for the SLE systems has also been reduced by 11% when comparing the predictions from this work ($AAD_{SLE} = 0.89$) with the predictions from our previous work[24] ($AAD_{SLE} = 1.00$). This improvement can be observed in Figure 9.

Analogous to the observations from the LLE calculations, the improvements observed in the predictions for SLE systems are driven by the improvements of the short-range and long-range terms. However, given that in this case salts are being modelled at the solubility limit, concentrations are higher and ion pairing related phenomena are more probable than in LLE systems where the salts tend to remain in the salt-rich phase. Consequently, the ion pairing related adjustments have a considerable dominance in reducing the overall deviation of the SLE predictions.

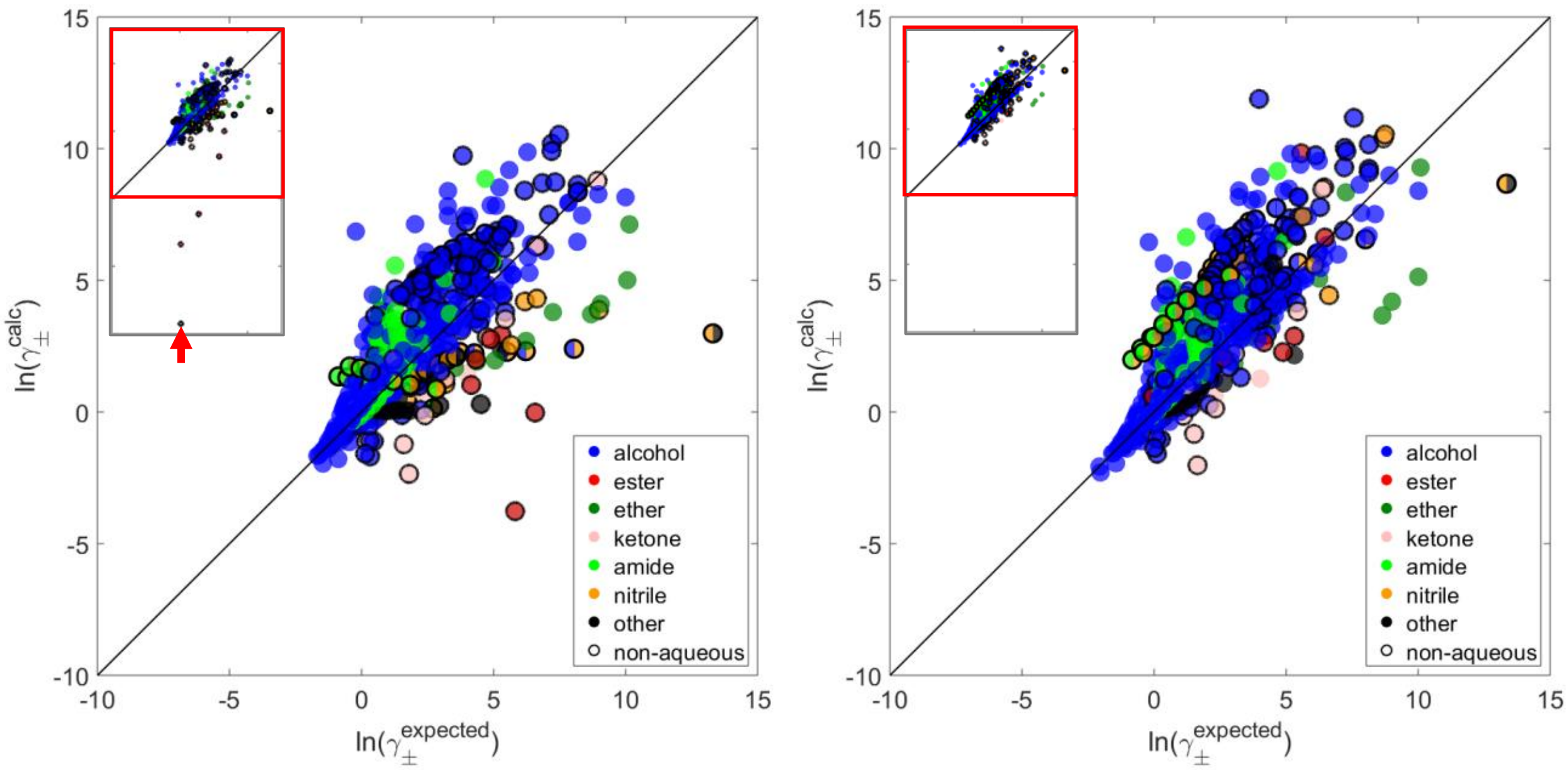

**Figure 9. Predicted activity coefficients vs. expected activity coefficients at 25°C. Solvents are grouped by functional group and in solvents with more than one functional group, other groups take priority over the alcohol group. Non-aqueous systems with two organic solvents**

**are shown with two colors. Left: predictions with the parameters from Müller et al[24]. Right: this work.**

Several outliers can be observed in Figure 9 (left), these outliers correspond to salts with a high solubility in ethyl acetate ($\epsilon_S \approx 6$). The most extreme case, where the logarithm of the calculated activity coefficient has an underestimated value close to -30, is the system $LiClO_4$ + diethyl ether ($\epsilon_S \approx 4.3$) system[48] (pointed by a red arrow). The other ester systems that are not outliers correspond to salts that have a very low solubility (i.e. $RbClO_4$) in this solvent[48]. As shown in Figure 9 (right), ion pairing considerations have led to correcting all of these cases.

Salts like $LiClO_4$ have a high solubility in diethyl ether and ethyl acetate (around 50 wt%[48]) and evidence suggests that in diethyl ether the free ions are absent and the salt is solvated as contact ion pairs and ion aggregates[49]. As a consequence, long-range Coulombic contributions would seem mild if not negligible. In our model this is accounted for by a very large association constant that makes the effective ionic strength tend to small values. Hence, in this case the calculated activity coefficient in equation (13) is practically given only by the COSMO-RS based short-range term for the salt since the influence of the long-range term becomes negligible.

The effective ionic strength in these low permittivity systems is considerably smaller when compared to the ionic strength assuming full dissociation. The ion concentration is inversely proportional to the squared Debye length[50]. One interpretation of the results of our model would be that the electrostatic decay length of the long-range term is increased at higher salt concentrations because the number of free ions is reduced in favor of neutral ion pairs. This in turn diminishes the contribution of the long-range term. While ion pairing in our case could be interpreted as an increase of the Debye length, it remains unclear whether this supports any type of causality or explanation of the physical phenomena observed in experimental studies[50]. Whether the structure of highly concentrated electrolytes is predominantly given by ion pairing effects or not is a matter of debate[51] and it is clearly beyond the scope of the present work to address such issues. From the chemical engineering modelling perspective it can only be pointed out that, for these highly concentrated systems, applying the Coulombic contribution (PDH term) only for dissociated ions resulted in systematic qualitative improvements and this is in agreement with recent observations reported in the literature[41,42,50].

Figure 10 shows solubility predictions for several salts in different mixed solvents as an example of calculations done by the new version of the model. In most cases the trends of the different systems have been improved. Analogous to LLE systems, the model is very accurate for mixtures of a salt in water-alcohol mixtures. This is not surprising, given that alcohol containing systems predominate in our database.

SLE systems are used to test the predictive capabilities of the model given that these are concentrated systems and many data points contain pure organic solvents. It may be observed in Figure 10 that in absolute terms all predictions are further improved and some are now remarkably accurate (i.e. $RbNO_3$ in water-methanol mixtures). In the case of polar aprotic solvents the trend for the NaCl + water + DMSO system[52] shown in Figure 10 (right) and the solubility of NaBr in acetonitrile[53] show improvements in the present work. However, further analysis is required for the adequate handling of these types of solvents. This observation agrees with the deviations found for ketone containing LLE systems.

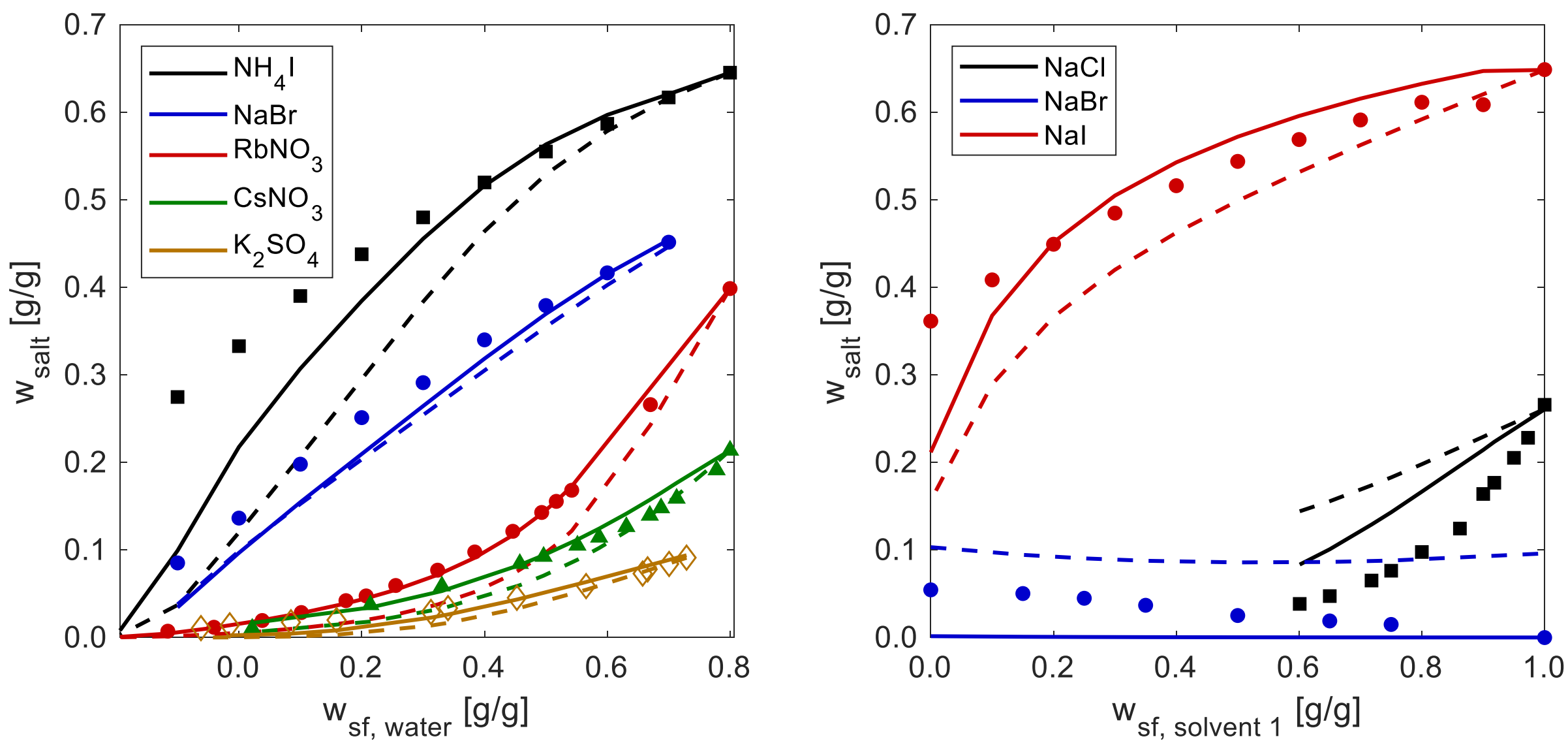

**Figure 10. Solubility predictions at 25°C from COSMO-RS-ES plotted as function of the salt free concentration of solvent (1). Markers correspond to the experimental values, dashed lines correspond to calculations performed with our previous parameterization[24] and solid lines correspond to this work. Left: $NH_4I$ + water(1) + ethanol(2)[54]; NaBr + water(1) + ethanol(2)[55]; + $RbNO_3$ + water(1) + methanol(2)[56]; $CsNO_3$ + water(1) + tert-butanol(2)[57]; $K_2SO_4$ + water(1) + chloral-hydrate(2)[58]. Right: NaI + water(1) + formamide(2)[59]; NaCl + water(1) + DMSO(2)[52]; NaBr + acetonitrile(1) + DMSO(2)[53].**

In Figure 10 the trends are correctly reproduced and the absolute values are predicted quite accurately. Even for these systems with strong non-ideal behavior at high salt concentrations, the complexity of the ion-organic solvent interactions in diverse non-aqueous systems is described by COSMO-RS-ES with only a few parameters. The ammonium iodide system[54] in Figure 10 (left) is a good example of one of the areas of opportunity for even further improvements: a salt with high solubility can be accurately described in a mixed aqueous alcohol solvent; but as the water fraction becomes smaller, the prediction deviates more from the experimental values. The description of a highly soluble salt in non-aqueous media is in many cases still challenging. An analogous observation can be made for the solubility of NaBr in DMSO in Figure 10 (right).

Future work could require the inclusion of systems with high salt concentrations or activity and/or osmotic coefficients of salts in non-aqueous solvents in the training set. Were these included in the training set, we would expect a better description of concentrated non-aqueous electrolytes. Even though this type of data is not so commonly reported in the literature, it may present an opportunity to improve the model. It is also relevant to mention that the effect that ion pairs and free ions have on the static permittivity of the solution have not yet been considered.

Overall, salt solubility predictions in mixed solvents have been made more accurate by the aforementioned modifications of the COSMO-RS-ES model while keeping the same number of parameters. This reaffirms the potential of the model to correlate and also predict a wide variety of electrolyte systems and proves that additional physical considerations can strengthen its predictive power.

## 5. Conclusions

The present work demonstrates that the predictive power of the COSMO-RS-ES model for the calculation of electrolyte systems can be further improved through the consideration of ion pairing without the need for additional non-general parameters. For this purpose, a Bjerrum treatment-based ion pairing method was introduced into the long-range term to adjust the ionic strength of the solution. As a result, the long-range contribution is modulated as a function of the concentration of free ions and the static permittivity of the solvent. Such considerations are combined with simple modifications to the empirical short range ion interaction energy equations and lead to an overall improvement of the LLE calculations by 25% and of the SLE predictions by 11%.

These changes have a negligible effect on the evaluations of Gibbs free energies of transfer of ions and aqueous mean ionic activity coefficients and do not decrease the accuracy to calculate these systems when compared to previous COSMO-RS-versions.

Overall, the LLE calculations have been improved for almost all systems. Major improvements driven by the modified short-range term equations were observed for acetonitrile containing systems. Large improvements driven by the ion pairing considerations were observed for systems with a low static permittivity in one of the liquid phases. Similar observations can be made for SLE systems, where all the most extreme outliers from our previous work[24] are now corrected.

Both the predictions and calculations still have some deviations. The most relevant ones are: firstly, an underestimation of the affinity of salts for completely non-aqueous systems that contain polar aprotic solvents. Secondly, a systematic underestimation of the solubility of highly soluble salts in pure organic solvents. This in turn presents additional opportunities for further improvement in the near future.

In general, COSMO-RS-ES has proven to be a versatile $g^E$-model for the description of electrolyte systems and the present work demonstrates there are still opportunities for improvement for this predictive model without the need for additional parameters. This reinforces its adequacy to study systems with limited experimental data reported in the literature and potentially aid in decision making processes like the selection of solvents or inorganic salts for specific applications.

**Declarations of interest:** none